\documentclass[runningheads]{llncs}
\usepackage[T1]{fontenc}
\usepackage{graphicx}
\usepackage{amsmath}

\makeatletter
\newif\if@anonymize

\@anonymizefalse

\begin{document}
\sloppy
\title{Noise2Contrast: Multi-Contrast Fusion Enables Self-Supervised Tomographic Image Denoising}
\titlerunning{Multi-Contrast Fusion Denoising}
%
\if@anonymize
\author{No authors given}
\authorrunning{No authors given}
\institute{No institute given\\
\email{No email given}}
\else
\author{Fabian Wagner\inst{1} \and
Mareike Thies\inst{1} \and
Laura Pfaff\inst{1, 2} \and
Noah Maul\inst{1, 2} \and
Sabrina Pechmann\inst{3} \and
Mingxuan Gu\inst{1} \and
Jonas Utz\inst{4} \and
Oliver Aust\inst{5} \and
Daniela Weidner\inst{5} \and
Georgiana Neag\inst{5} \and
Stefan Uderhardt\inst{5} \and
Jang-Hwan Choi\inst{6} \and
Andreas Maier\inst{1}}
\authorrunning{F. Wagner et al.}
\institute{Pattern Recognition Lab, FAU Erlangen-Nürnberg, Germany \and
Siemens Healthcare GmbH, Erlangen, Germany \and
Fraunhofer Institute for Ceramic Technologies and Systems IKTS, Germany \and
Department AIBE, FAU Erlangen-Nürnberg, Germany \and
Department of Rheumatology and Immunology, FAU Erlangen-Nürnberg, Germany \and
Division of Mechanical and Biomedical Engineering, Ewha Womans University, Korea\\
\email{fabian.wagner@fau.de}}
\fi
\maketitle              
\begin{abstract}
Self-supervised image denoising techniques emerged as convenient methods that allow training denoising models without requiring ground-truth noise-free data. Existing methods usually optimize loss metrics that are calculated from multiple noisy realizations of similar images, e.g., from neighboring tomographic slices. However, those approaches fail to utilize the multiple contrasts that are routinely acquired in medical imaging modalities like MRI or dual-energy CT. In this work, we propose the new self-supervised training scheme Noise2Contrast that combines information from multiple measured image contrasts to train a denoising model. We stack denoising with domain-transfer operators to utilize the independent noise realizations of different image contrasts to derive a self-supervised loss. The trained denoising operator achieves convincing quantitative and qualitative results, outperforming state-of-the-art self-supervised methods by $4.7$--$11.0\,\%$/$4.8$--$7.3\,\%$ (PSNR/SSIM) on brain MRI data and by $43.6$--$50.5\,\%$/$57.1$--$77.1\,\%$ (PSNR/SSIM) on dual-energy CT X-ray microscopy data with respect to the noisy baseline. Our experiments on different real measured data sets indicate that Noise2Contrast training generalizes to other multi-contrast imaging modalities.

\keywords{Self-Supervised Denoising \and Known Operator Learning \and Contrast Fusion.}
\end{abstract}
\section{Introduction}
\label{sec:intro}
Measured data is inherently affected by uncertainty determined by the measurement process and its related physics. In image data, that uncertainty appears as image noise, disturbing an underlying ground truth image signal. Whereas imaging parameters like acquisition time, detector sensitivity, or illumination can be chosen to keep noise levels low, realistic imaging settings usually require a trade-off between acquisition parameters and image quality. In fact, some measurements, e.g., in clinical workflows, can only be carried out by accepting severe amounts of noise due to radiation exposure, acquisition time, or patient motion. Therefore, image processing algorithms were developed to reduce noise levels and extract the underlying noise-free signal. Conventional algorithms robustly denoise image data but require expert knowledge to adapt the algorithm to domain-specific conditions \cite{tomasi1998bilateral}. Unlike conventional filters, learning-based models can learn task-specific features purely from a training data distribution without domain-specific knowledge. However, deep neural networks inherently lack interpretability and were shown to be prone to prediction artifacts on out-of-domain samples \cite{wagner2022trainable}. Different hybrid approaches tried to combine data-driven optimization with conventional image filters to create reliable denoising operators with close to state-of-the-art performance \cite{wagner2022ultra}.

\begin{figure}[tb]
\centering
\includegraphics[width=\textwidth]{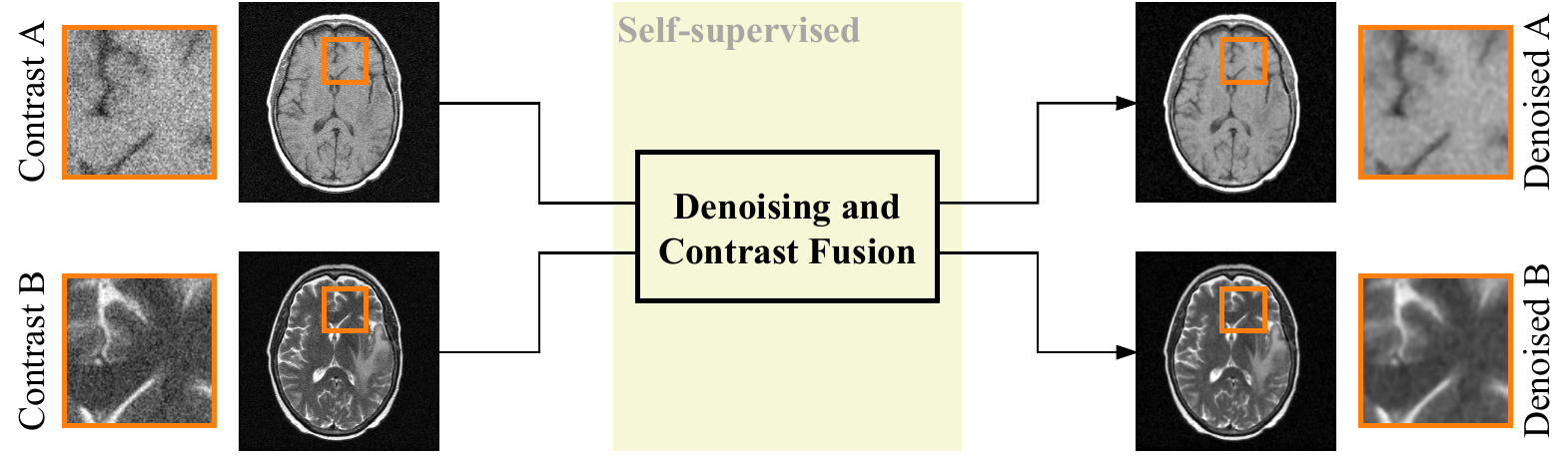}
\caption{Noise2Contrast: Fusion of image contrasts A and B enables self-supervised denoising, e.g., using T1 and T2 weighted MRI scans.} \label{fig:title_fig}
\end{figure}

Recently, multiple self-supervised denoising methods were proposed, circumventing the need for ground truth noise-free data during training \cite{lehtinen2018n2n,krull2019noise2void,batson2019noise2self,kim2022noise}. Noise2Noise \cite{lehtinen2018n2n} and Noise2Void \cite{krull2019noise2void} allow self-supervised image denoising using two noisy representations of the same image or pixel-wise masking to calculate loss metrics that do not require a ground truth. Different other works applied these concepts to medical imaging modalities, e.g., by using neighboring volumetric slices \cite{zhang2021noise2context,jeon2022mm} or time frames \cite{wu2020self} as training targets following the Noise2Noise scheme.

Although self-supervised training on individual medical scans showed promising results, most existing approaches are not capable of using all available data. Many used medical imaging modalities like Magnetic Resonance Imaging (MRI) or dual-energy Computed Tomography (DECT) routinely acquire multiple image contrasts of the same scanned object that remain so far unused in self-supervised denoising approaches. In this work, we present the novel denoising method Noise2Contrast which is capable of using multiple image contrasts to train a denoising model in a fully self-supervised manner. An overview of Noise2Contrast is illustrated in Fig.\,\ref{fig:title_fig}. Our method is able to employ the independent noise realizations in different image contrasts of medical imaging modalities to train a robust denoising operator. We confirm our theoretical considerations with extensive experiments on real medical data. Our contributions are three-fold.
\begin{itemize}
  \item We present the self-supervised denoising method Noise2Contrast combining image information from different acquired image contrasts.
  \item We demonstrate how to train a robust denoising operator using our proposed scheme by simultaneously learning denoising and domain transformation.
  \item Extensive experiments quantitatively and qualitatively confirm the applicability of our method on different real medical data sets.
\end{itemize}

\section{Methods}
\label{sec:methods}

\subsection{Self-Supervised Image Denoising}
\label{ssec:selfsup}
Each image acquisition $j$ introduces noise $n$ through the image formation and detection processes to the ground truth object $y$
\begin{align}
    x_i^{(j)} = y_i + n_j\enspace.
\end{align}
Image denoising then aims to find an operator $f_w$ that maps noise-affected images $x_i^{(j)}$ to a denoised prediction $\hat{y}_i$ close to the noise-free ground truth $y_i$ by minimizing
\begin{align}
    \underset{w}{\text{argmin}} \sum_i \mathcal{L} \left(f_w\left(x_i^{(1)}\right), y_i\right)
    \label{eq:sup_optimiz}
\end{align}
based on a loss metric $\mathcal{L}$ and parameters $w$. Supervised learning methods typically use a training set of $N$ paired samples $\left(x_i^{(1)}, y_i\right)$ with $i \in \{1, \dots, N\}$ to train a neural network data driven to predict denoised images from the learned training data distribution. As paired ground truth images are often difficult to obtain in real applications, self-supervised training methods aim to find an optimal denoising operator while having solely access to noisy images. Lehtinen et al.~\cite{lehtinen2018n2n} demonstrated that learning the mapping of the noisy measurement to a second image with the same content but a different noise realization $x_i^{(2)}$, e.g., a second photo taken, is similar to solving the supervised problem in Eq.\,\ref{eq:sup_optimiz}
\begin{align}
    \underset{w}{\text{argmin}} \sum_i \mathcal{L} \left(f_w\left(x_i^{(1)}\right), x_i^{(2)}\right)\enspace.
    \label{eq:selfsup_optimiz}
\end{align}
Although many works adopt this so-called Noise2Noise training scheme, the method requires at least two images with equivalent content and contrast per sample during training which might not be available in reality. Other works, e.g., Noise2Void \cite{krull2019noise2void}, propose masking individual pixels of noisy images to create pseudo-paired training samples $x_i^{(1\star)}$. Subsequently, a denoising model can be trained by learning to predict the correct intensity values at the masked positions. However, Noise2Void demands pixel-wise statistically independent noise which is often not satisfied in particular on real detector data and for medical imaging modalities \cite{wagner2022ultra}.

\subsection{Denoising Using Known Operators}
\label{ssec:knownoperator}
Including prior knowledge in neural network architectures has been shown beneficial in terms of model performance, generalizability, and prediction robustness \cite{maier2019learning,wagner2022trainable}. We adapt the known operator learning concept in our proposed method by separating denoising and domain-transfer tasks through the network architecture as described in section \ref{ssec:mcfusion}. As the denoising operator, we use trainable bilateral filter layers \cite{wagner2022ultra} that can be trained via gradient-based optimization like any other neural network layer. The filter forward operation smooths image content in homogeneous regions (spatial kernel) while preserving edges through a range kernel
\begin{align}
    \hat{Y}_k &= \frac{1}{\alpha_k} \sum_{n \in \mathcal{N}} G_{\sigma_s}(\left\lVert k - n \right\rVert) G_{\sigma_r}(X_k - X_n) X_n
    \label{eq:defBF}
\end{align}
with
\begin{align}
    \alpha_k &= \sum_{n \in \mathcal{N}} G_{\sigma_s}(\left\lVert k - n \right\rVert) G_{\sigma_r}(X_k - X_n)
\end{align}
and $G_{\sigma_s}$ and $G_{\sigma_r}$ denoting Gaussian spatial and range kernel of width $\sigma_s$ and $\sigma_r$ respectively. $\left\lVert \dots \right\rVert$ indicates the spatial distance between pixels of index $k$ and $n$ and $\mathcal{N}$ is the filter window. The differentiable implementation of Wagner et al.~\cite{wagner2022ultra} allows optimizing all filter parameters $\sigma_s$ and $\sigma_r$ data driven using deep learning frameworks. The algorithmic filter design from Eq.\,\ref{eq:defBF} proves that the bilateral filter can solely act as a denoising operator as it is not able to extract complex features or modify the images besides local pixel intensity averaging.

In addition to data-driven optimization of a known denoising algorithm, we demonstrate how to employ a neural network as an independent denoising operator. By training denoising and domain-transfer networks subsequently, different image processing tasks can be entirely separated into independent network parts to enable self-supervised denoising of multi-contrast data. The proposed training schemes are presented in the following section.

\begin{figure}[tbh]
\includegraphics[width=\textwidth]{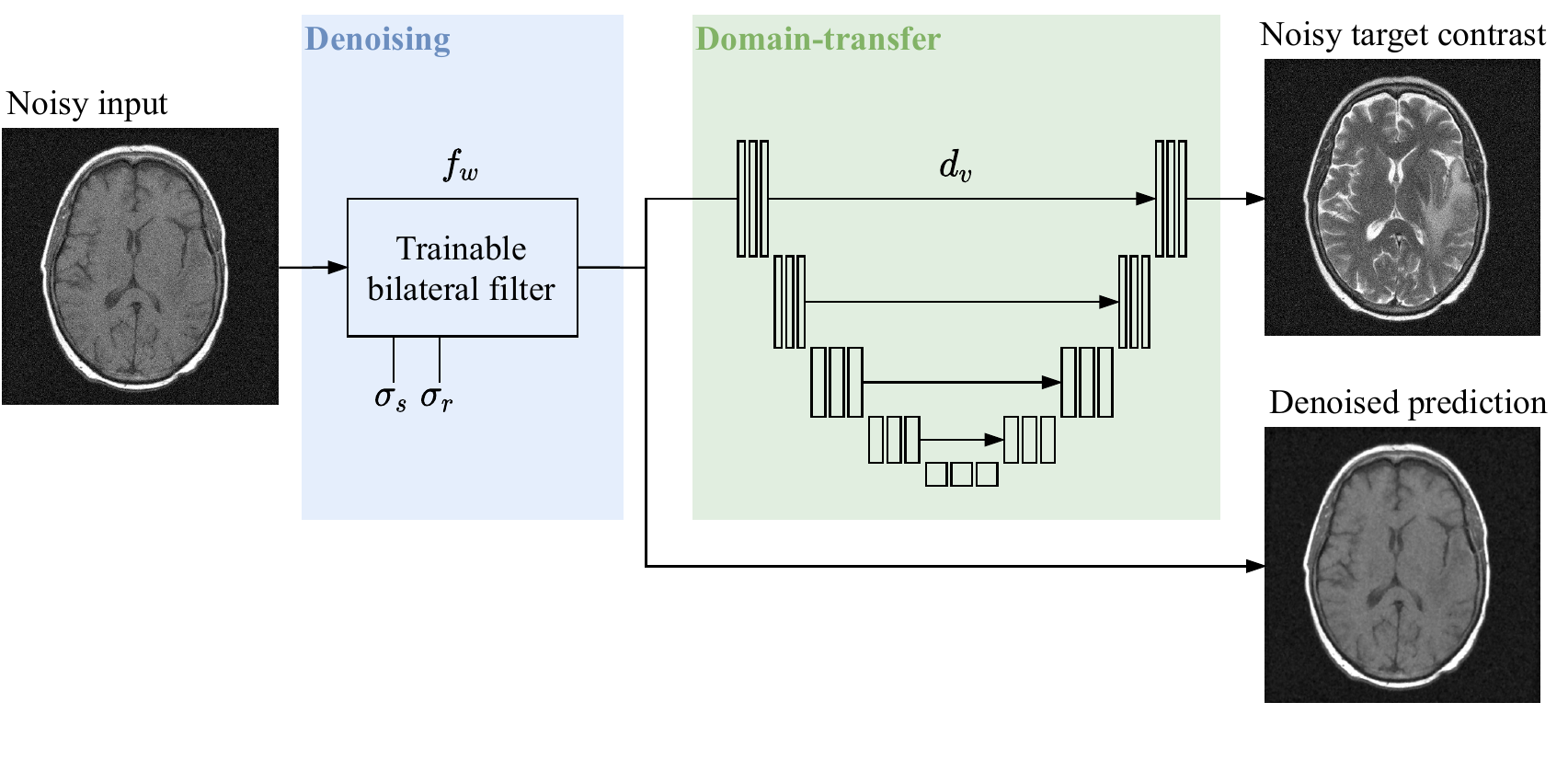}
\caption{Illustration of the proposed domain-transfer-based self-supervised denoising approach Noise2Contrast on the example of MRI T1 and T2-weighted contrasts. A noisy input of contrast one is processed by subsequent denoising (blue) and domain-transfer (green) operators. This allows deriving a self-supervised loss metric $\mathcal{L}$ using the noisy target with contrast two. The denoised input image is obtained by removing the domain-transfer operator.} \label{fig:pipeline}
\end{figure}

\subsection{Multi-Contrast Fusion Through Domain Transfer}
\label{ssec:mcfusion}
Measuring two images with the same content to perform Noise2Noise denoising is often infeasible in medical imaging due to radiation and time constraints. However, modalities like MRI or DECT routinely acquire multiple image contrasts that show the same anatomical structures but highlight different biological features. A second noisy image contrast indicated by $\dag$ (e.g., T1 and T2 weighting in MRI imaging)
\begin{align}
    x_i^{\dag(j)} = y_i^{\dag(j)} + n
\end{align}
could be used as a noise-affected target image in the setting of Eq.\,\ref{eq:selfsup_optimiz}. In such a setting, a network would learn to predict a denoised image with the target contrast. However, it is not possible to extract a solely denoised image $\hat{y}_i$ from the network prediction with preserved contrast. To avoid mixing both tasks, we propose separating the trained model into known operators to allow using the network parts individually during inference as we are only interested in the denoised prediction but want to preserve the original image contrast. An illustration of the presented training scheme is illustrated in Fig.\,\ref{fig:pipeline}. We present two solutions how to separate the denoising and domain translation tasks to enable self-supervised denoising.

\subsubsection{Known Operator-based} First, a known denoising operator is used in combination with a domain translation neural network $d_v$ and trained self supervised. We use a trainable bilateral filter layer (Sec. \ref{ssec:knownoperator}) as the filter operation can not perform complex domain translations or intensity shifts by design and thus can be considered as a known denoising operator. Therefore, denoising and domain translation are inherently separated through the pipeline's architecture when training the chained operators $d_v$ and $f_w$. The following training task results
\begin{align}
    \underset{w, v}{\text{argmin}} \sum_i \mathcal{L} \left(d_v\left(f_w\left(x_i^{(1)}\right)\right), x_i^{\dag(1)}\right)
    \label{eq:selfsup_mcfusion}
\end{align}
containing the domain translation operator $d_v$ represented by a U-Net \cite{ronneberger2015u} with trainable parameters $v$. A self-supervised mean squared error loss is calculated between the denoised and domain-translated input image and the target contrast image $x_i^{\dag(1)}$ with independent noise.

\subsubsection{Network Operator-based} Second, a neural network is trained as a denoising operator in the same setting as Eq.\,\ref{eq:selfsup_mcfusion}. To enforce a strict separation of denoising and domain translation, operators $d_v$ and $f_w$ are trained in a subsequent fashion. First, the domain translation network is trained in the known operator-based setting to predict images of target contrast $y_i^{\dag}$ from denoised input contrast images $\hat{y}_i$. Subsequently, that trained network is frozen and employed as a domain translation operator to transfer the predictions of a denoising neural network to the target contrast domain. The sequential training of denoising and domain translation operator enforces the networks to learn tasks independently and use them as separate image processing operators.

\section{Experiments}
\label{sec:experiments}

\subsection{Data}
\label{ssec:data}
We perform multiple experiments to investigate how noise can be effectively reduced in multi-contrast medical data without requiring noise-free ground truth data. First, we evaluate our method on three different MRI contrasts that are routinely used to identify tissue-specific properties and abnormalities: T1, T2, and Fluid Attenuated Inversion Recovery (FLAIR)-weighting. We use the public Brain-Tumor-Progression data set \cite{schmainda2018btp} consisting of clinical MRI head scans of 20 brain tumor patients and split it into twelve training, two validation, and six test patients. Each scan contains T1, T2, and FLAIR-weighted reconstructions that are used as input and target data to evaluate the proposed self-supervised denoising method. We simulate Gaussian noise as present in the real and imaginary part of complex-valued reconstructed MR images or in the phase-corrected magnitude images \cite{prah2010simple} and choose the noise standard deviation as $5\,\%$ of the maximum scan intensity.

In a second experiment, we compare denoising methods on a mouse tibia bone sample scanned in a dual-energy Zeiss Xradia 620 Versa X-ray Microscope (XRM). Tomographic XRM imaging is instructive for investigating bone-remodeling and bone-related diseases on the micrometer scale due to its high bone-to-soft tissue contrast. Here, dual-energy acquisitions allow quantitative measurements of bone density and sample composition \cite{genant1977quantitative}. However, dual-energy XRM measurements contain severe noise levels due to finite scan times and dose concerns in potential in vivo measurements \cite{wagner2022monte}. We denoise a $1.5\,\text{h}$ dual-energy scan ($50\,\text{kV}$ and $70\,\text{kV}$) and compare the predictions with a $14\,\text{h}$ high-SNR acquisition that is regarded as ground truth. XRM scans are reconstructed using the pipeline of Thies et al.~\cite{thies2022calibration}. The two settings LE (low-energy) $\rightarrow$ HE (high-energy) and $\text{HE}\rightarrow\text{LE}$ are investigated.

\subsection{Networks}
\label{ssec:networks}
Three stacked trainable bilateral filter layers \cite{wagner2022ultra} are employed as known operator-based denoising model $f_w$. The domain translation network $d_v$ is represented by a standard U-Net \cite{ronneberger2015u} with $16$ input features and around $1.1\,\text{Mio}$ trainable parameters $v$. We use the Adam optimizer with learning rate $5\cdot 10^{-5}$ in all our experiments. Models are trained until convergence of the self-supervised training loss computed on the validation scans in each epoch (MRI data) or on the training scan (XRM data).

\subsection{Denoising Experiments}
\label{ssec:denoising_experiments}
Different contrast combinations are investigated for the MRI data set to evaluate the generalizability of our proposed self-supervised denoising approach. We chose the settings $\text{T1}\rightarrow\text{T2}$, $\text{T2}\rightarrow\text{T1}$, and $\text{T2}\rightarrow\text{FLAIR}$ for our experiments with the respective input $x_i$ and target $x_i^{\dag}$ contrast domains $\text{input}\rightarrow\text{target}$.

We compare our methods to the state-of-the-art blind-spot training scheme Noise2Void (N2V) \cite{krull2019noise2void}. In addition, we compare to a different reference method where a target image $x_i^{(2)}$ is chosen as the neighboring slice of the input image $x_i^{(1)}$. Multiple related works apply this or similar principles to create pseudo-pairs of noisy images \cite{zhang2021noise2context,jeon2022mm,choi2022self}. We denote the reference approach as Noise2Neighbor\,(N2N) in the following as it comes close to the initial Noise2Noise idea where two noisy images of the same contrast are available. Our known operator and network operator-based methods are denoted as Noise2Contrast\,(BFs) and Noise2Contrast\,(U-Net) respectively.

\section{Results}
\label{sec:results}

\begin{table}
\caption{Quantitative denoising results on the Brain-Tumor-Progression \cite{schmainda2018btp} MRI test data set. (mean $\pm$ std) is calculated over the patients. The best-performing method is highlighted in bold.}\label{tab:btp_results}
\begin{tabular}{llcc}
\hline
Setting & Method & PSNR (mean $\pm$ std) \hphantom{I} & SSIM (mean $\pm$ std)\\
\hline
& Noisy baseline & $26.02 \pm 0.01$ & $0.384 \pm 0.059$ \\
& Noise2Contrast\,(BFs) & $\mathbf{36.76 \pm 1.40}$ & $\mathbf{0.869 \pm 0.021}$ \\
$\text{T1}\rightarrow\text{T2}$ & Noise2Contrast\,(U-Net) & $30.43 \pm 0.20$ & $0.385 \pm 0.119$ \\
& Noise2Void (BFs) \cite{krull2019noise2void} & $35.81 \pm 1.56$ & $0.847 \pm 0.023$ \\
& Noise2Neighbor\,(BFs) \cite{zhang2021noise2context,choi2022self} & $32.49 \pm 4.36$ & $0.867 \pm 0.047$ \\
\hline
& Noisy baseline & $26.02 \pm 0.02$ & $0.444 \pm 0.071$ \\
& Noise2Contrast\,(BFs) & $\mathbf{34.69 \pm 1.91}$ & $\mathbf{0.865 \pm 0.023}$ \\
$\text{T2}\rightarrow\text{T1}$ & Noise2Contrast\,(U-Net) & $33.19 \pm 1.03$ & $0.698 \pm 0.070$ \\
& Noise2Void (BFs) \cite{krull2019noise2void} & $34.30 \pm 1.69$ & $0.841 \pm 0.024$ \\
& Noise2Neighbor\,(BFs) \cite{zhang2021noise2context,choi2022self} & $29.91 \pm 3.80$ & $0.828 \pm 0.067$ \\
\hline
& Noisy baseline & $26.02 \pm 0.02$ & $0.444 \pm 0.071$ \\
& Noise2Contrast\,(BFs) & $\mathbf{35.21 \pm 1.70}$ & $\mathbf{0.871 \pm 0.022}$ \\
$\text{T2}\rightarrow\text{FLAIR}$ & Noise2Contrast\,(U-Net) & $21.74 \pm 0.05$ & $0.221 \pm 0.135$ \\
& Noise2Void (BFs) \cite{krull2019noise2void} & $34.30 \pm 1.69$ & $0.842 \pm 0.023$ \\
& Noise2Neighbor\,(BFs) \cite{zhang2021noise2context,choi2022self} & $29.90 \pm 3.80$ & $0.828 \pm 0.067$ \\
\hline
\end{tabular}
\end{table}

\begin{table}
\caption{Quantitative denoising results on the dual-energy XRM bone scan. (mean $\pm$ std) is calculated over the z-slices. The best-performing method is highlighted in bold.}\label{tab:dexrm_results}
\begin{tabular}{llcc}
\hline
Setting & Method & PSNR (mean $\pm$ std) \hphantom{A} & SSIM (mean $\pm$ std)\\
\hline
& Noisy baseline & $22.17 \pm 0.29$ & $0.158 \pm 0.008$ \\
$\text{LE}\rightarrow\text{HE}$ \hphantom{AA} & Noise2Contrast\,(BFs) & $\mathbf{29.86 \pm 0.19}$ & $\mathbf{0.622 \pm 0.015}$ \\
& Noise2Void (BFs) \cite{krull2019noise2void} & $27.28 \pm 0.22$ & $0.420 \pm 0.015$ \\
\hline
& Noisy baseline & $23.15 \pm 0.29$ & $0.178 \pm 0.010$ \\
$\text{HE}\rightarrow\text{LE}$ \hphantom{AA} & Noise2Contrast\,(BFs) & $\mathbf{30.63 \pm 0.22}$ & $\mathbf{0.610 \pm 0.015}$ \\
& Noise2Void (BFs) \cite{krull2019noise2void} & $28.36 \pm 0.24$ & $0.453 \pm 0.015$ \\
\hline
\end{tabular}
\end{table}

\begin{figure}[!htb]
\begin{center}
\includegraphics[width=\textwidth]{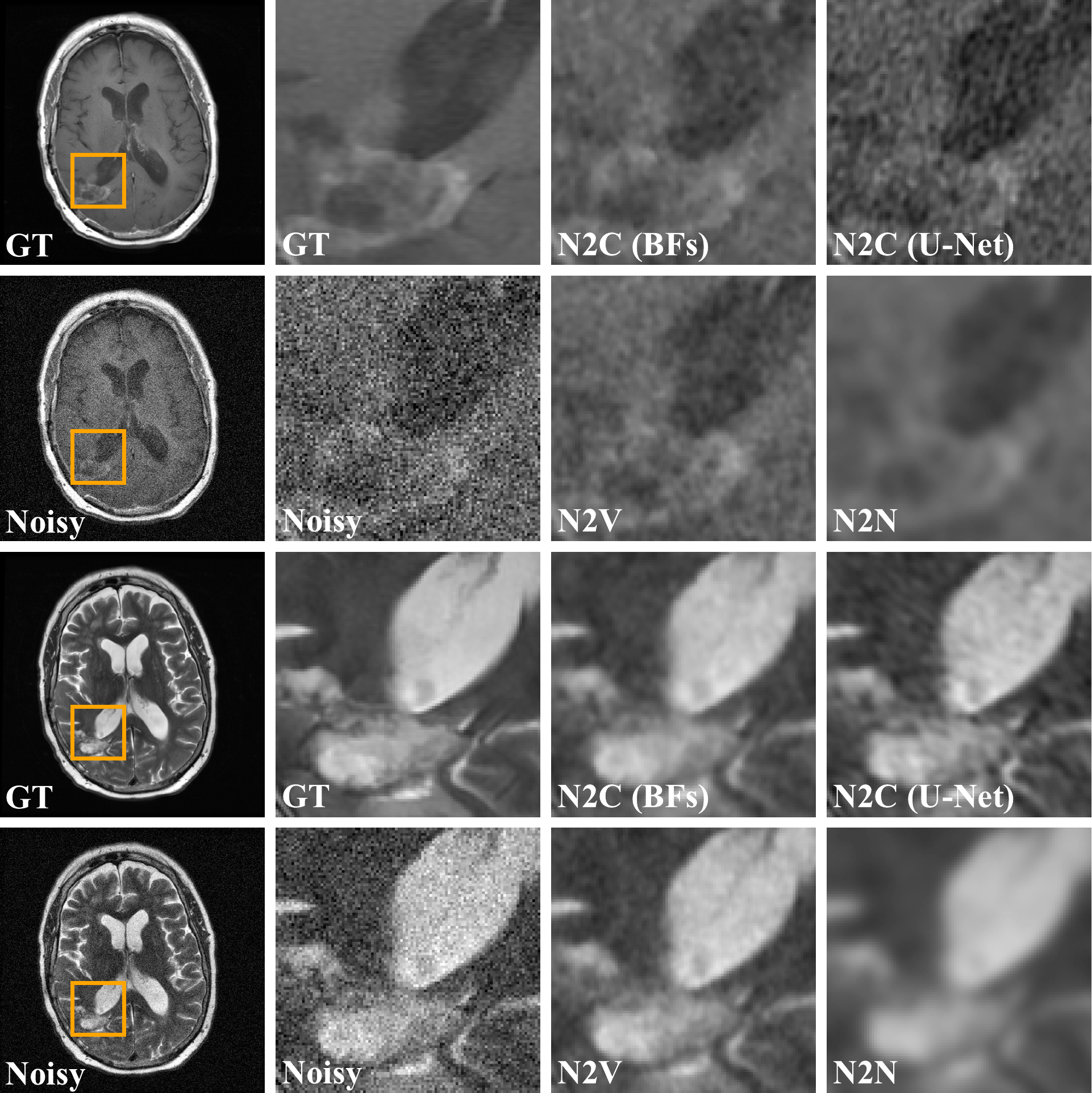}
\caption{Qualitative denoising results on the Brain-Tumor-Progression \cite{schmainda2018btp} MRI test data set of $\text{T1}\rightarrow\text{T2}$ (top) and $\text{T2}\rightarrow\text{T1}$ predictions. The images are displayed in equal windows.} \label{fig:btp_rois}
\end{center}
\end{figure}

\begin{figure}[!htb]
\begin{center}
\includegraphics[width=\textwidth]{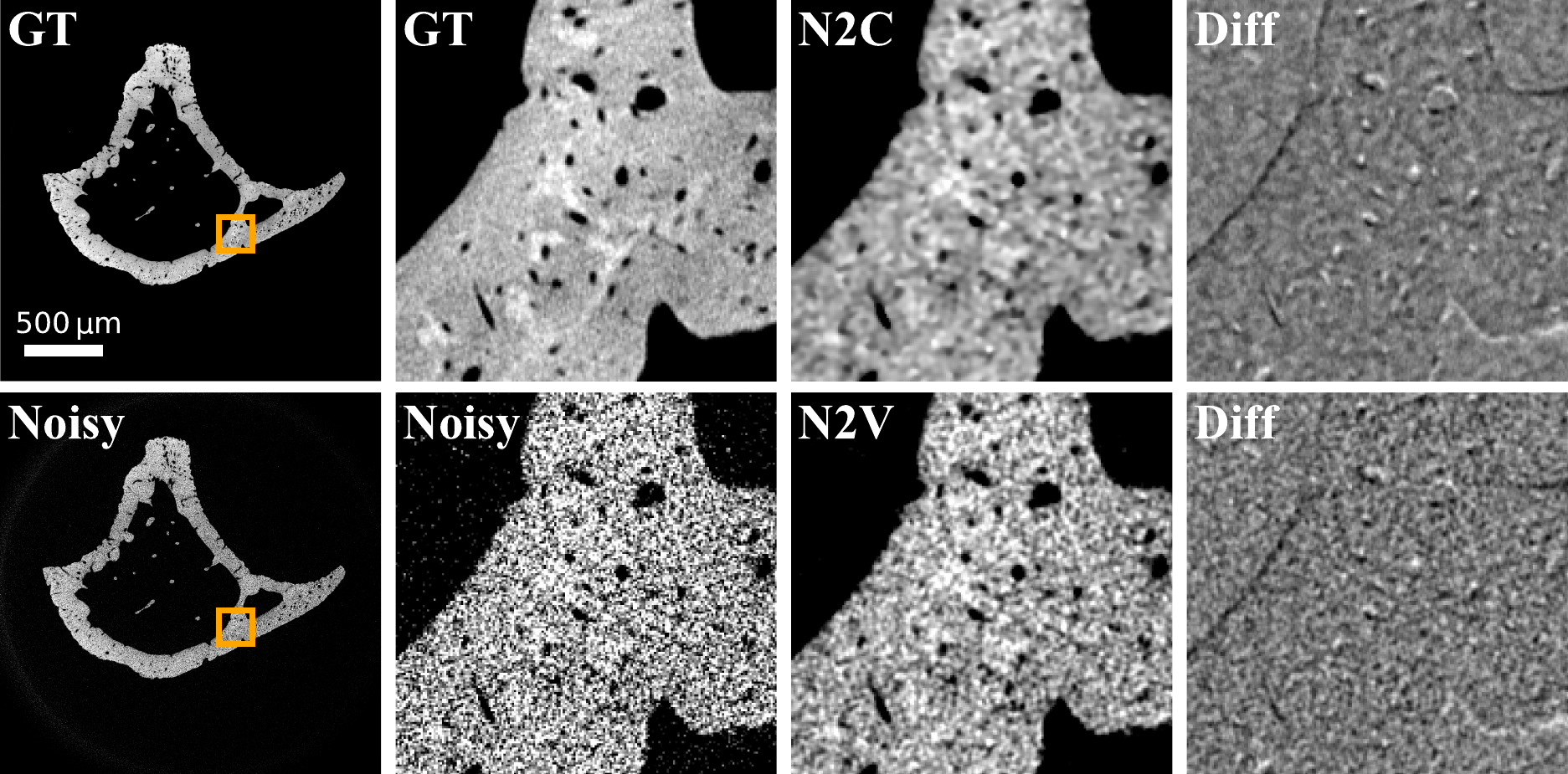}
\caption{Qualitative denoising results on the dual-energy XRM bone scan in the $\text{LE}\rightarrow\text{HE}$ setting. The images are displayed in equal windows. Diff denotes the difference images between the respective method and the high-dose ground truth.} \label{fig:dexrm_rois}
\end{center}
\end{figure}

We compute the quantitative image quality metrics peak signal-to-noise ratio (PSNR) and structural similarity index measure (SSIM) for all model predictions. Results on the Brain-Tumor-Progression data are presented in Tab.\,\ref{tab:btp_results}. Our proposed multi-contrast training scheme using known operators Noise2Contrast\,(BFs) quantitatively outperforms all comparison methods. Noise2Contrast\,(BFs) improves the results of Noise2Void by $4.7$--$11.0\,\%$ PSNR and by $4.8$--$7.3\,\%$ SSIM with respect to the noisy baseline. Exemplary predictions visualized in Fig.\,\ref{fig:btp_rois} confirm the quantitative findings and show that our training scheme converges to a solution that preserves features while removing the image noise. Predictions of our additional experiment using a U-Net for denoising (Noise2Contrast\,(U-Net)) exhibit lower noise removal compared to the known operator-based method. State-of-the-art Noise2Void training achieves similar visual results compared to our method, however, predictions contain a slightly higher noise level. Noise2Neighbor fails to predict reasonable images and blurs high-frequency features. The lower half of the magnified regions in Fig.\,\ref{fig:btp_rois} contains a brain lesion that allows comparing perceptual noise levels on a clinical pathology.

Results on the dual-energy XRM data are presented in Tab.\,\ref{tab:dexrm_results} and Fig.\,\ref{fig:dexrm_rois}. Here, Noise2Contrast\,(BFs) improves the results of Noise2Void by $43.6$--$50.5\,\%$ PSNR and by $57.1$--$77.1\,\%$ SSIM with respect to the noisy baseline. On par with the quantitative metrics, the visual predictions of Noise2Void contain considerably more noise than our presented Noise2Contrast\,(BFs) training. This is particularly visible in the provided difference images that are calculated between the model predictions and the $14\,\text{h}$ high-SNR XRM acquisitions. Note that the low-dose network input and the high-dose ground-truth scans are independently acquired scans. Despite the high mechanical precision of the used XRM, subsequent scanning results in small micrometer-scale shifts that are visible as thin edges in the difference images.

\section{Discussion}
\label{sec:discussion}

The training configuration using a U-Net-based denoising model in combination with a domain-transfer model Noise2Contrast\,(U-Net) achieved promising visual results. However, the quantitative performance left room for improvement compared to the best-performing methods. We recognized that the trained denoising U-Net predicted visually appealing results but did not always fully preserve all input contrast intensities which led to poor quantitative metrics. We believe that a better-designed and more thoroughly trained domain-transfer model would help to provide more reasonable image gradients to the denoising network and improve the overall denoising performance. Pre-trained domain transfer models that are trained on ground truth data \cite{denck2021enhanced} can be employed here to improve the domain transfer operation. Alternatively, a regularizing loss term calculated between denoised input contrast and noisy input image can be investigated to enforce preserved intensities.

We performed additional experiments directly mapping the input contrast to the target contrast image with a single model following the standard Noise2Noise approach. Although such models learned to simultaneously denoise and map to the target domain, their clinical application remains very limited as the model predictions inherently alter the image contrast which is usually not desired. In this Noise2Noise setting, image quality metrics calculated between model prediction and input contrast ground truth yielded poor scores as expected due to the modified prediction contrast. Additionally, we investigated a setting with a known denoising operator like the trainable bilateral filter used to predict the denoised input contrast by mapping on the target contrast without using a domain translation network. This yielded poor results likewise as the known denoising operator is not capable of learning the contrast mapping such that it only predicted blurred images to minimize the training loss.

Only a few fully self-supervised denoising techniques exist that can remove noise while preserving high-frequency image features. Blind-spot methods like Noise2Void can achieve impressive results on certain data sets but are limited to pixel-wise independent noise statistics by design. Therefore, compelling results can be achieved on imaging modalities with simple noise characteristics and simulated data like the Brain-Tumor-Progression MRI scans in the first part of our study. Real measured data and computed tomography scans generally contain correlated noise caused by the detection process and the image reconstruction algorithm. Our experiments on real measured dual-energy XRM data confirm this limitation of Noise2Void. In contrast, our proposed known operator-based training scheme Noise2Contrast achieves considerably better quantitative and qualitative results as it does not depend on particular noise properties in the measured and reconstructed image data. Therefore, we conclude that Noise2Contrast is better suited to train models on modalities with correlated noise patterns like dual-energy CT compared to state-of-the-art Noise2Void training.

\section{Conclusion}
\label{sec:conclusion}

In this work, we presented the Noise2Contrast training scheme that allows self-supervised image denoising using multi-contrast data. Noise2Contrast combines information from independently measured image contrasts through an operator-based pipeline to train a denoising model. Our experiments on routine clinical MRI contrasts and on a pre-clinical dual-energy tomographic X-ray Microscope bone scan demonstrate superior performance of Noise2Contrast compared to the few other existing self-supervised denoising techniques. We believe that the universal Noise2Contrast training scheme can be applied on data from many more multi-contrast imaging modalities like photon-counting-CT, confocal microscopy, or hyperspectral imaging.

\if@anonymize
\else
\subsubsection{Acknowledgements} This work was supported by the European Research Council (ERC Grant No. 810316) and a GPU donation through the NVIDIA Hardware Grant Program.
\fi

%
%
%
\bibliographystyle{splncs04}
\bibliography{refs}
\end{document}